\documentclass[times, review, 10pt]{elsarticle}

\usepackage{amssymb}
\usepackage{amsmath}
\DeclareMathOperator*{\argmax}{arg\,max} 
\usepackage[linesnumbered,ruled,vlined]{algorithm2e}
\usepackage{float}
\usepackage{bbding}
\usepackage{booktabs}
\usepackage{multirow}
\usepackage{hyperref}
\hypersetup{
    pdfauthor={Runmin Jiang},
    colorlinks=true,
    linkcolor=blue,
    filecolor=magenta,      
    urlcolor=cyan,
}
\usepackage[capitalize]{cleveref}
\Crefname{table}{Tab.}{Tab.}
\crefname{figure}{Fig.}{Fig.}
\Crefname{equation}{Eq.}{Eq.}
\Crefname{algorithm}{Algorithm\ }{Algorithm\ }
\Crefname{appendix}{Appendix\ }{AppenDICEs\ }
\usepackage{microtype}
\newcommand{\etal}{\textit{et al.}}
\newcommand{\ie}{\textit{i.e.},}

\begin{document}

\begin{frontmatter}

\title{Enhancing Weakly Supervised 3D Medical Image Segmentation through Probabilistic-aware Learning}

\author[cmu]{Runmin Jiang}
\author[cmu]{Zhaoxin Fan\corref{equal}}
\author[towson]{Junhao Wu}
\author[cmu]{Lenghan Zhu} 
\author[towson]{Xin Huang}
\author[uab]{Tianyang Wang}
\author[umd]{Heng Huang}
\author[cmu]{Min Xu\corref{equal}}

\cortext[equal]{Corresponding authors: Min Xu\ead{mxu1@cs.cmu.edu}, Zhaoxin Fan}

\affiliation[cmu]{organization={Carnegie Mellon University},
            city={Pittsburgh},
            postcode={15213},
            state={PA},
            country={USA}}
            
\affiliation[towson]{organization={Towson University},
            city={Towson},
            postcode={21252},
            state={MD},
            country={USA}}
            
\affiliation[uab]{organization={University of Alabama at Birmingham},
            city={Birmingham},
            postcode={35294},
            state={AL},
            country={USA}}
            
\affiliation[umd]{organization={University of Maryland},
            city={College Park},
            postcode={20742},
            state={MD},
            country={USA}}

\begin{abstract}
3D medical image segmentation is a challenging task with crucial implications for disease diagnosis and treatment planning. Recent advances in deep learning have significantly enhanced fully supervised medical image segmentation. However, this approach heavily relies on labor-intensive and time-consuming fully annotated ground-truth labels, particularly for 3D volumes. To overcome this limitation, we propose a novel probabilistic-aware weakly supervised learning pipeline, specifically designed for 3D medical imaging. Our pipeline integrates three innovative components: a Probability-based Pseudo Label Generation technique for synthesizing dense segmentation masks from sparse annotations, a Probabilistic Multi-head Self-Attention network for robust feature extraction within our Probabilistic Transformer Network, and a Probability-informed Segmentation Loss Function to enhance training with annotation confidence. Demonstrating significant advances, our approach not only rivals the performance of fully supervised methods but also surpasses existing weakly supervised methods in CT and MRI datasets, achieving up to 18.1\% improvement in Dice scores for certain organs. The code is available at \href{https://github.com/runminjiang/PW4MedSeg}{https://github.com/runminjiang/PW4MedSeg}.
\end{abstract}

\end{frontmatter}

\section{Introduction}
\label{sec:intro}

Medical image segmentation is pivotal in refining healthcare systems for accurate disease diagnosis and strategic treatment planning, as it delineates anatomical structures across various imaging modalities, providing crucial information for healthcare professionals \cite{chen2021transunet}. Deep learning techniques have significantly impacted this field, evidenced by advancements in traditional supervised learning methods, particularly in 2D or 3D `U-shaped' encoder-decoder architectures like U-Net \cite{ronneberger2015u, zhang2021transfuse, milletari2016v, wang2022uctransnet, cao2021swin}. Despite their wide usage, these methods often require intensive manual annotation, a process that can be both time-consuming and resource-intensive \cite{TAJBAKHSH2020101693}. To mitigate these challenges, researchers have explored various strategies such as data augmentation \cite{zhang2021understanding, panfilov2019improving, fu2018three}, transfer learning \cite{ma2019neural, qin2019transfer}, and domain adaptation \cite{huo2018adversarial, chen2019unsupervised} to reduce reliance on extensive labeled data.

Nevertheless, weakly supervised training methods, employing minimal annotations like points and scribbles for generating pseudo labels, have gained increasing attention \cite{li2022fully, bearman2016s, lin2016scribblesup}. These approaches, while addressing the issue of manual annotation, predominantly focus on 2D image segmentation and often overlook the complexities of 3D weak annotation. This oversight can lead to significant information loss, as these methods tend to directly use sparse weak annotations during training. Furthermore, the confidence level of the annotator is frequently disregarded, omitting a vital aspect of the segmentation process.

In response to these challenges, we propose a novel weakly supervised pipeline for 3D medical image segmentation, emphasizing probability integration throughout training and inference. Inspired by the uncertainty model~\cite{gal2016dropout}, our approach transforms sparse 3D point labels into dense annotations through Probability-based Pseudo Label Generation. We further introduce a Probabilistic Multi-head Self-Attention mechanism within our Probabilistic Transformer Network to address class variance and noise in pseudo labels. Complementing this is our Probability-informed Segmentation Loss Function, which incorporates annotation confidence, aligning closer with true segmentation boundaries. This holistic approach, encompassing pseudo label generation, network structure, and loss function, effectively utilizes dense weakly supervised signals and reduces bias in confidence allocation, facilitating efficient segmentation with minimal annotation costs.

Solid experiments conducted on the authoritative BTCV and CHAOS datasets, representing CT and MRI images respectively, demonstrate the substantial efficacy of our approach. Our method consistently delivers exceptional results on both datasets, with noteworthy improvements -- achieving up to an 18.1\% and 10.2\% boost in Dice scores compared to point-supervised methods, as well as remarkable enhancements of 58.4\% and 17.6\% over scribble-supervised methods. Importantly, our method achieves results similar to or even surpasses one of the fully supervised tests. Further, we conducted dedicated ablation experiments on our framework's three critical components, encompassing pseudo label generation, network structure, and loss function. Remarkably, all these components yielded positive results, collectively contributing to the enhanced accuracy of segmentation within our framework. These findings underscore our method's potential as a robust and versatile solution for medical image segmentation in weakly supervised settings. 

The main contributions of our approach can be summarized as follows:
\begin{itemize}
\item \textbf{Probabilistic-aware Framework}: We introduce a novel probabilistic-aware weakly supervised learning pipeline. Through a comprehensive series of tests, we demonstrate that our method not only significantly enhances performance compared to state-of-the-art weakly supervised methods but also achieves results comparable to fully supervised approaches, highlighting its substantial real-world applicability.

\item \textbf{Probability-based Pseudo Label Generation}: Within the framework, we innovate by converting sparse 3D point labels into comprehensive dense annotations, leveraging principles from the uncertainty model. This innovative approach minimizes the typical information loss associated with weak labels and enhances segmentation accuracy. Additionally, we simulated the diversity of real-world raw data to test the practicality of our method and achieved promising results.

\item \textbf{Probabilistic Multi-head Self-Attention (PMSA)}: A critical component of our probabilistic transformer network, it effectively addresses the inherent class variance and noise found in pseudo labels. It plays a pivotal role in enhancing segmentation performance by capturing and utilizing the probabilistic distributions of input-output mappings.
    
\item \textbf{Probability-informed Segmentation Loss Function}: To complement the framework, we introduce a novel loss function that incorporates the annotator's confidence level. This loss function aligns the segmentation process more closely with actual boundaries and captures the probabilistic nature of the segmentation task. It also plays a crucial role in reducing the bias in confidence allocation during model training.
\end{itemize}

\section{Related Work}

\subsection{Medical Image Segmentation}
This task is dedicated to extracting objects of interest from medical images obtained through modalities such as Computed Tomography (CT) and Magnetic Resonance Imaging (MRI). Fully Convolutional Networks (FCN)~\cite{long2015fully} and U-Net~\cite{ronneberger2015u} have significantly advanced 2D medical image segmentation. Adjustments to U-Net by Guan \etal~\cite{guan2019fully} and Ibtehaz \etal~\cite{ibtehaz2020multiresunet} have been put forward to enhance the precision of segmentation. For 3D volumetric medical image segmentation, Cicek \etal~\cite{cciccek20163d} introduces a 3D U-Net that handles spatial information from 2D slices, while Milletari \etal~\cite{milletari2016v} presents V-Net with improved feature extraction and reduced computational costs. However, the primarily discussed techniques are fully supervised methods tailored for 2D medical image segmentation. In contrast, our paper emphasizes weakly supervised approaches for 3D medical image segmentation, aiming for more efficient annotation processes.

\subsection{Weakly Supervised Segmentation}
Weakly supervised learning reduces annotation cost by using sparse annotations instead of fully annotated masks. Weak labels such as bounding boxes~\cite{rother2004grabcut, dai2015boxsup}, scribbles~\cite{lin2016scribblesup}, and points~\cite{bearman2016s} have been utilized. Zhang \etal~\cite{zhang2021interactive} integrates point-level annotation and sequential patch learning for CT segmentation. Roth \etal~\cite{roth2021going} designs a point-based loss function with an attention mechanism. Zou \etal~\cite{zou2020pseudoseg} proposes a well-calibrated pseudo-labeling strategy, while Liu \etal~\cite{liu2022acpl} introduces an informative selection strategy. In contrast, our work proposes a "dense" weak annotation approach from a probabilistic perspective.

\subsection{Probabilistic Modeling in Deep Learning}
Probabilistic modeling in deep learning handles uncertainty and provides confidence intervals. Shirakawa \etal~\cite{shirakawa2018dynamic} uses a Bernoulli distribution to generate network structures. Choi \etal~\cite{choi2021active} estimates a probabilistic distribution using mixture density networks for object detection. Zhang \etal~\cite{zhang2021bayesian} introduces Bayesian attention belief networks, while Guo \etal~\cite{guo2022uncertainty} scales dot-product attention as Gaussian distributions. Our method is the first probabilistic modeling approach for 3D medical image segmentation, incorporating probability in annotation, network structure, and gradient backpropagation, offering advantages for training and inference.

\begin{figure*}[ht]
        \centering
        \includegraphics[width=\linewidth]{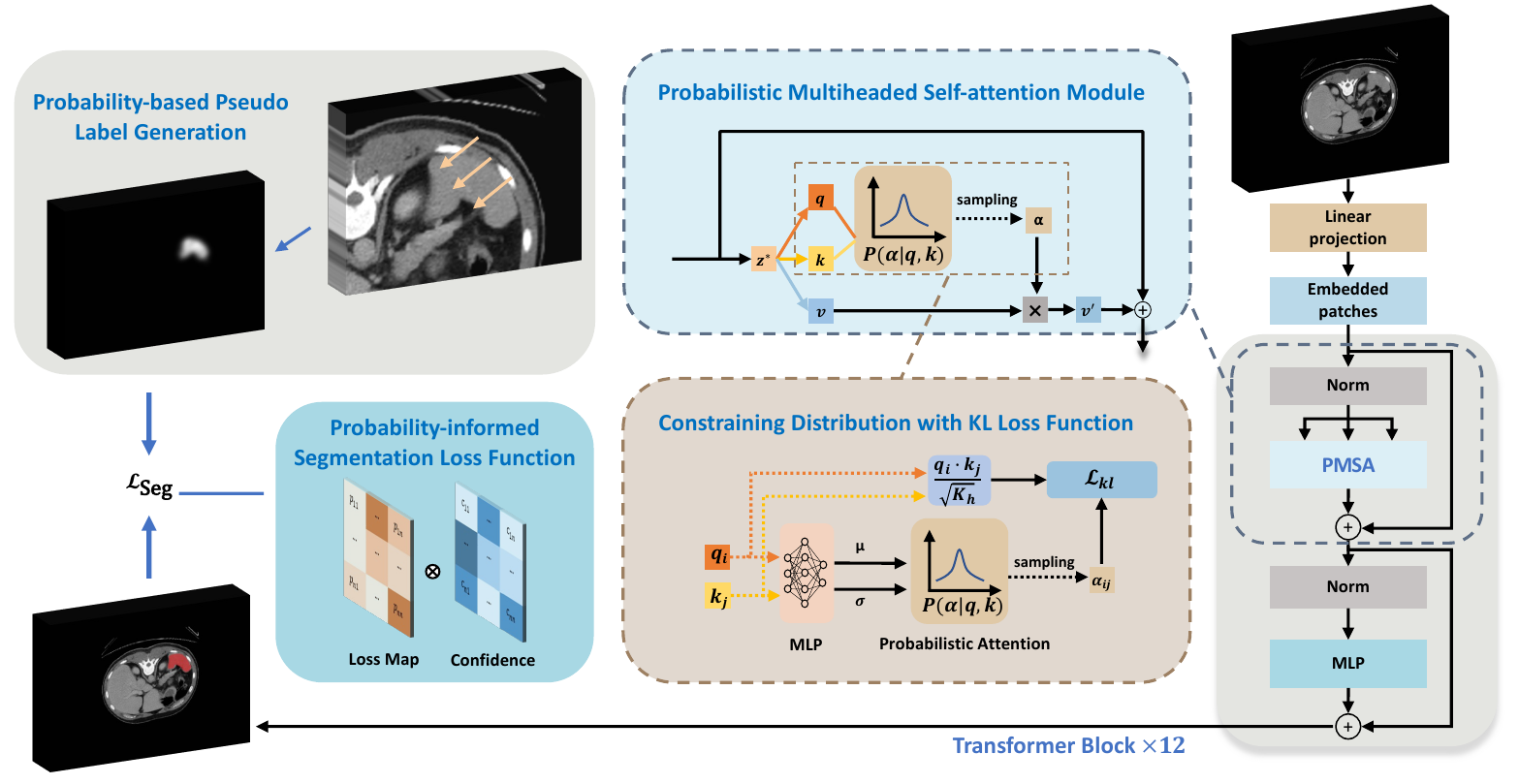}
        \vspace{-0.3cm}
        \caption{\textbf{Overview of our framework.} We adopt UNETR \cite{hatamizadeh2022unetr} as the baseline network for our segmentation model. The input is a 3D medical volume, which is processed by our Probabilistic Transformer Network, which is powered by the PMSA mechanism. The output of the network is a 3D segmentation map, which is supervised by the ``dense" probability-based pseudo label generated from ``sparse" point-based annotations. A Probability-aware Segmentation Loss Function is proposed to train the network.} 
         \vspace{-0.05in}
        \label{fig:overview}
\end{figure*}

\section{Method}
\subsection{Overview}
Medical image segmentation, which is typically referred to as semantic segmentation of medical images, aims to partition the image into different non-overlapping regions with unique semantic labels. Given an image $I$ and the semantic classes $\{C_{1},C_{2},...,C_{k}\}$, the semantic segmentation process is performed by dividing $I$ into $\{D_{1},D_{2},...,D_{k}\}$ (\ie~the subregions), which satisfies:
\begin{equation}
\label{eq0}
 I=\bigcup\limits_{i=1}^{k}{D_{i}}, \quad D_{i}\cap{D_{j}}=\emptyset, \quad \forall{i\neq{j}},\; i,j\in{k}
\end{equation}
where k is a positive integer no less than 2, and all pixels in the region $D_{i}$ are labeled with $C_{i}(i=1,2,...,k)$. In a weak supervision setting, the model is trained on a training set denoted as $\{(I_{1},L_{1}^{*}),(I_{2},L_{2}^{*}),...,(I_{n},L_{n}^{*})\}$, where $I_{i}(i=1,2,...,n)$ is the image and $L^{*}_{i}(i=1,2,...,n)$ is the weak label. During inference, the model outputs dense segmentation of the input images.

\cref{fig:overview} illustrates the overview of our method for solving the weakly supervised 3D medical image segmentation task. We introduce a novel weakly supervised training pipeline for 3D medical image segmentation, taking probabilistic features of both annotation process and network training into consideration. We illustrate our pipeline in the following aspects: 1) A probability-based pseudo label generation scheme for generating ``dense" weak annotations. 2) A probabilistic Transformer network, whose key component is the proposed gaussian-based multi-head self-attention mechanism. 3) The probability-informed loss function. 

\subsection{Probability-based Pseudo Label Generation}
\label{ppmg}
\subsubsection{Sparse Labels Annotation}
\label{annotation strategy}
In this paper, we explore weakly supervised 3D medical image segmentation to lower annotation costs, choosing 3D points for sparse labeling. This approach helps in generating high-quality pseudo dense labels by instructing annotators to select random, evenly distributed 3D points on the organ's surface. Experimentally, we simulate this process by eroding the ground-truth label with a $5\times5\times5$ structuring element and then applying Farthest Point Sampling (FPS) to pick points within this eroded region. This method ensures an even distribution of points, effectively mimicking real annotation and creating pseudo sparse labels that closely represent the organ's surface distribution.

\begin{figure}[htbp]
  \centering
  \includegraphics[width=0.9\linewidth]{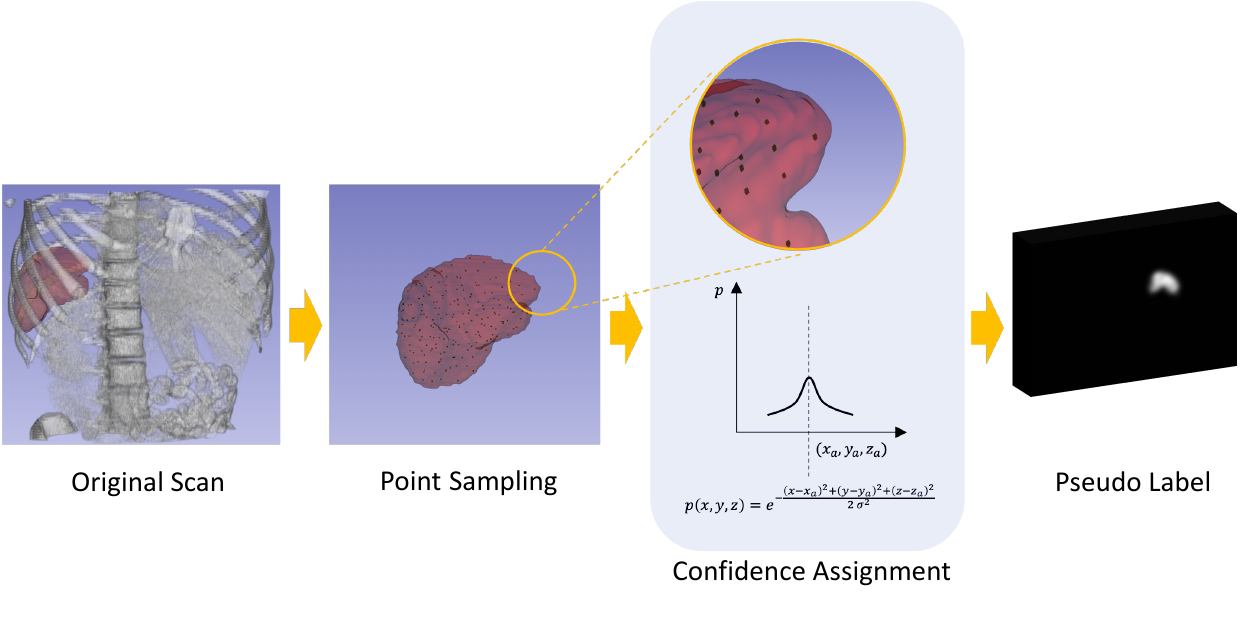}
  \vspace{-0.2in}
  \caption{\textbf{The pipeline for Probability-based Pseudo Label Generation.} The points are randomly sampled within the target organ. The Probability-based Pseudo Label is generated by assigning confidence using the sampled points.}
  \label{pseudo label generation}
  \vspace{-0.1in}
\end{figure}

\subsubsection{Pseudo Label Generation}
\label{dense annotation}
After acquiring sparse labels, directly using them for supervision leads to substantial information loss and may inadequately train a 3D medical image segmentation network. To overcome this, we introduce a method for generating dense 3D labels. This method is based on the idea that annotated points and their vicinity possess confidence scores, decreasing with distance from the point. Specifically, for an annotated point $(x_a, y_a, z_a)$, we apply a Gaussian function to model confidence scores, peaking at the annotated point and diminishing with distance. The confidence score $P(x, y, z)$ for any point $(x, y, z)$ is defined as:

\begin{equation}
P(x, y, z) = e^{-\frac{(x - x_a)^2 + (y - y_a)^2 + (z - z_a)^2}{2\sigma^2}}
\end{equation}

This process, applied to all annotated points, generates dense 3D labels by summing up the label maps from all points and normalizing the intensity to $[0, 1]$. This probability-based pseudo label generation scheme effectively transforms sparse annotations into informative dense labels, improving the training of the segmentation network. The entire Probability-based Pseudo Label Generation pipeline is illustrated in \cref{pseudo label generation}, while the algorithm of target generation is described in \cref{alg1}. 

The core idea of this algorithm is to generate a three-dimensional Gaussian distribution based on the coordinates of each sampled point and to accumulate these distributions to form the final label map. This label map can subsequently be used for training medical image segmentation models. The variance \(\sigma^2\) influences the width of the generated Gaussian distribution, thereby altering the shape of the label map.

\begin{algorithm}[tb]
    \caption{Target Generation}
    \label{alg1} 
    \LinesNumbered
    \KwIn{$C=\{x_i,y_i,z_i\}_{i=1}^{n}$: coordinates of sampled points, $X$, $Y$, $Z$}
    \KwOut{$P\in{\mathbb{R}^{X\times{Y}\times{Z}}}$: label map}
        \For{$\{x_i,y_i,z_i\}$ in $C$}{
            $P_{i}\gets{0}$\;
    	\For{$x\gets{0}$ to $X$}{
    		\For{$y\gets{0}$ to $Y$}{
    		      \For{$z\gets{0}$ to $Z$}{
    		          $P_i(x,y,z)=e^{-\frac{(x-x_i)^2+(y-y_i)^2+(z-z_i)^2}{2\sigma^2}}$\;
    	        }
    	    }
    	}
        }
        $P=\sum_{i=1}^{n}P_i$\;
        Normalize $P$ to $[0,1]$\;   
\end{algorithm}

\subsection{Probabilistic Transformer Network}
Though the proposed pseudo label in \cref{dense annotation} can reflect the confidence level of the annotator, the within-class variance is high, due to the inherent morphological variation of human organs and the randomness of the point-sampling process. Therefore, a probabilistic model is expected to capture the complex distribution.

\subsubsection{Network Architecture}
\begin{figure}[tb]
        \centering
        \includegraphics[width=0.99\linewidth]{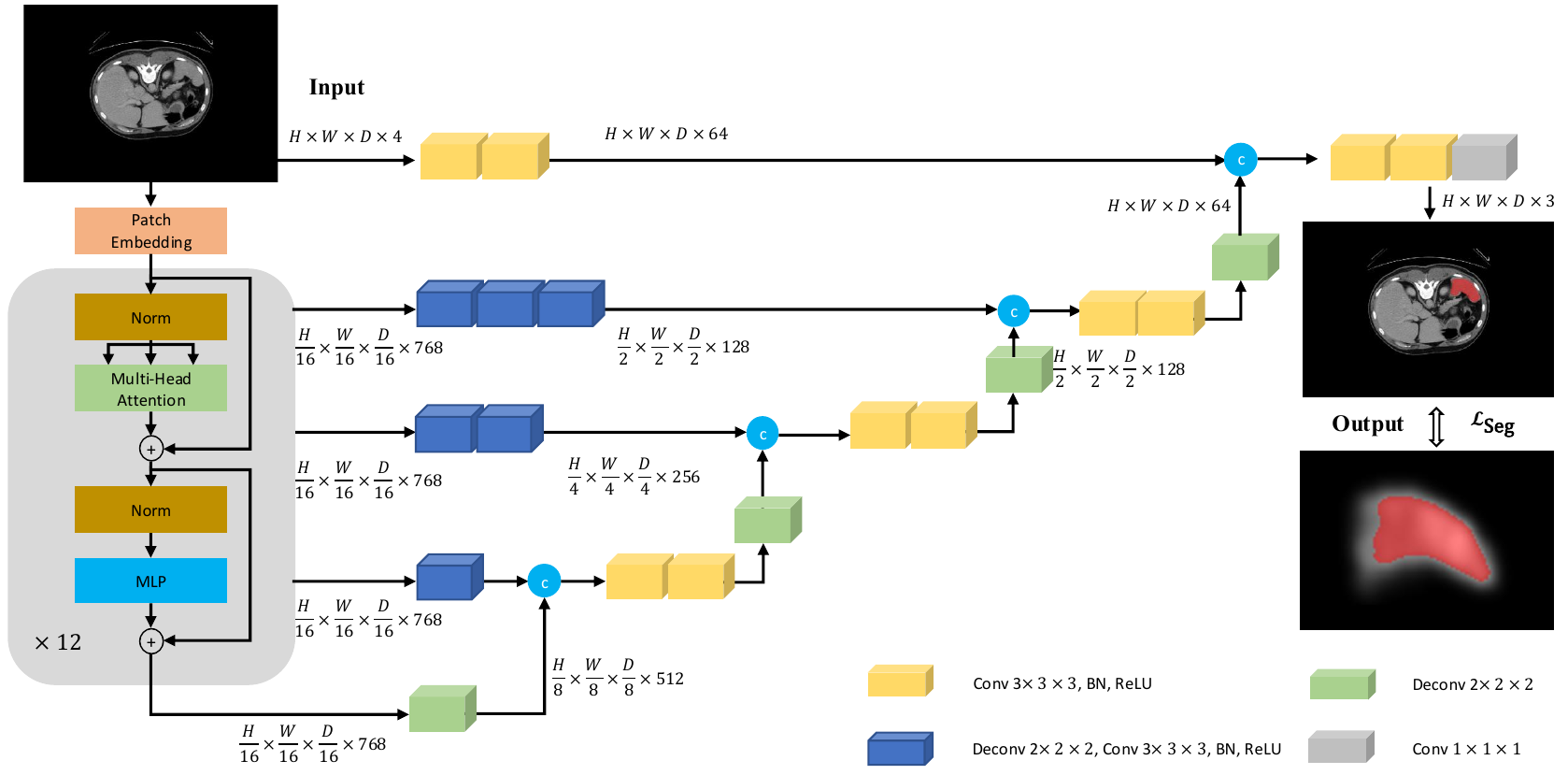}
        \vspace{-0.3cm}
        \caption{Network Architecture Overview} 
        \vspace{-0.05in}
        \label{network}
\end{figure}

Our framework adopts the contracting-expanding schema characteristic of the UNETR architecture, as illustrated in Fig.~\ref{network}. Initially, a 3D volume \(x \in \mathbb{R}^{H \times W \times D \times C}\), with dimensions \((H, W, D)\) and \(C\) input channels, is segmented into non-overlapping uniform patches of dimensions \((P, P, P)\). This segmentation transforms the volume into a sequence \(x_v \in \mathbb{R}^{N \times (P^3 \cdot C)}\) by flattening these patches, where \(N = (H \times W \times D) / P^3\) denotes the sequence length. Subsequently, these patches are mapped into a \(K\)-dimensional embedding space via a linear layer. Furthermore, a 1D learnable positional embedding \(E_{pos} \in \mathbb{R}^{N \times K}\) is incorporated into the mapped patches. The process can be defined as follows:

\begin{equation}
z_{0} = [x^{1}_{v}E; x^{2}_{v}E; ...; x^{N}_{v}E] + E_{pos}
\end{equation}

Here, $E \in \mathbb{R}^{(P^3 \cdot C) \times K}$ represents the patch embedding projection.
The features are then passed through a series of Probabilistic Transformer blocks, which consist of alternating layers of PMSA and MLP blocks. The equations for these blocks are as follows:

\begin{equation}
z_{i}' = \text{PMSA}(\text{Norm}(z_{i-1})) + z_{i-1}, \quad i = 1...L
\end{equation}

\begin{equation}
z_{i} = \text{MLP}(\text{Norm}(z_{i}')) + z_{i}', \quad i = 1...L
\end{equation}

\subsubsection{Probabilistic Multi-head Self-Attention}

Multi-head Self-Attention (MSA) is a key component in the Transformer model. It captures the dependencies between different positions in an input sequence by using multiple attention heads. In MSA, given an input sequence $z \in \mathbb{R}^{N \times K}$, where $N$ signifies the sequence length and $K$ signifies the feature dimension at each position, each attention head generates a set of attention weights to compute the attention values for each position concerning other positions. The calculation of MSA can be expressed as follows:

\begin{equation}
\text{MSA}(z) = \text{Softmax} \left( \frac{QK^T}{\sqrt{d_k}} \right) V
\end{equation}

Here, $Q$, $K$, and $V$ are obtained by linearly transforming the input sequence $z$ into query, key, and value representations, respectively. The attention weights are computed by taking the dot product of the query and key vectors, scaled by the square root of the key dimension $d_k$. The softmax function is applied to obtain the final attention weights. Finally, the attention values are computed by multiplying the attention weights with the value vectors.

However, the Probability-based Pseudo Label suffers from large in-class variance caused by the randomness of the point-sampling process and the inherent diversity of human organ structure. To guide our model to capture the variance within the proposed pseudo label and encode the input properly, inspired by Guo \etal~\cite{guo2022uncertainty}, we introduce our Probabilistic Multi-head Self-Attention module. In a single SA head, we assume that the dependency score $\alpha_{ij}$ follows a Gaussian distribution: $\alpha_{ij} \sim \mathcal{N}(\mu_{ij}, \sigma^{2}_{ij})$, where the mean $\mu_{ij}$ and the variance $\sigma^{2}_{ij}$ are calculated with $q_{i}$ and $k_{j}$ using a multilayer perceptron (MLP). In order to allow the parameters to be updated through backpropagation, we adopt reparameterization trick~\cite{NIPS2015_bc731692}:
\begin{equation}
\label{eq9} 
\alpha_{ij}= \mu_{ij} + \sigma_{ij}\epsilon,\quad \epsilon\sim \mathcal{N}(0, 1)
\end{equation}
where $\epsilon$ is a random variable that follows a standard normal distribution.
For other parameters in the model, we set them as deterministic, and denote them as $\Theta$.

We assume that the dependency scores within the same PMSA layer are independent of each other, while the dependency scores of deeper PMSA layer are dependent on those of former PMSA layers: 
\begin{equation}
\label{eq10}
\alpha_{l}\sim p(\alpha_{l}|X',\Theta,\alpha_{l-1}...,\alpha_{1}),\quad l=1,...,L
\end{equation} 
where $\alpha_{l}$ denotes the dependency scores of the PMSA layer of the $l$th transformer block. 

With PMSA, the distribution of the output segmentation map $y'$ given the input image $X'$ can be computed according to: 
\begin{equation}
\begin{split}
\label{eq11} 
P(y'|X', \Theta)=E_{\alpha \sim p(\alpha|X',\Theta)}[P(y'|X',\Theta,\alpha)] \\
= \int_{\alpha}{P(y'|X',\Theta,\alpha)p(\alpha|X',\Theta)d\alpha}
\end{split}
\end{equation}

However, due to the intractability of the integral in \cref{eq11}, we sample $\alpha$ from $p(\alpha|X',\Theta)$ for $M$ times to approximate the integral, in which every $\alpha_{ij}$ is sampled independently each time: 
\begin{equation}
\label{eq12}
y^{*}={\rm \mathop{argmax}\limits_{\it y'}}\sum\limits_{m=1}^{M}\frac{1}{M}P(y'|X',\Theta,\alpha_{m})
\end{equation}
where $\alpha_{m}$ denotes the dependency scores sampled at the $m$th time, and $y^{*}$ is the final segmentation output. More details about the sampling of dependency scores and the proof of \cref{eq11} can be found in ~\ref{app2}.

\subsection{Probability-informed Segmentation Loss Function}
As discussed in \cref{ppmg}, the proposed pseudo label is considered a probability map, where the intensity of each point represents the annotator's confidence in classifying it as the target organ. Therefore, to enable our model to be aware of the underlying confidence within the pseudo label, we introduce a loss function which is a combination of DICE loss and Probability-weighted Cross Entropy (PCE) loss. The intuition is that points with prior confidence greater than a certain threshold are considered as the foreground of the basic label map, while we weight the loss function with the prior confidence of the annotator since voxels with low confidence deserve lower loss weights.

Given the output \(y^*\) and pseudo label map \(S\), the segmentation loss is formulated as:

\begin{equation}
\mathcal{L}_{\text{Seg}} = \mathcal{L}_{\text{DICE}}(y^*, S_{T}) + \mathcal{L}_{\text{PCE}}(y^*, S_{T})
\end{equation}

where \(S_T\) is the thresholded map of \(S\) with a threshold \(T\) (set as 0.5), and for each voxel in the segmentation map, \(\mathcal{L}_{\text{pce}}\) is formulated as:

\begin{equation}
\mathcal{L}_{\text{pce}}(p_i, s_i) = \begin{cases}
s_i \log(p_i), & \text{if } s_i \geq T \\
(1 - s_i) \log(1 - p_i), & \text{if } s_i < T
\end{cases}
\end{equation}

where \(s_i\) and \(p_i\) are the confidence of the \(i\)th voxel in \(S\) and \(y^*\), respectively, and \(N\) denotes the number of voxels in the segmentation map. The PCE loss is then averaged over all voxels to obtain \(\mathcal{L}_{\text{PCE}}\):

\begin{equation}
\mathcal{L}_{\text{PCE}} = \frac{1}{N} \sum_{i=1}^{N} \mathcal{L}_{\text{pce}}(p_i, s_i)
\end{equation}

Moreover, to align the model's learned distribution of dependency scores with a realistic expectation of the data, we set our prior distributions based on empirical observations of the data and domain knowledge. The KL divergence loss is introduced to enforce this alignment:

\begin{equation}
\begin{aligned}
\label{eq13}
\mathcal{L}_{KL}= &\sum\limits_{l=1}^L{\sum\limits_{i,j}{\mathcal{D}_{KL}(p(\alpha_{lhij}|X',\Theta,\alpha_{k-1},  ..., \alpha_{1})|| 
\mathcal{N}(\alpha_{lhij}',\sigma^{2})))}} \\
= &\sum\limits_{l=1}^L{\sum\limits_{i,j}{\log\frac{\sigma}{\sigma_{lhij}}+\frac{\sigma_{lhij}^2+(\mu_{lhij}-\alpha_{lhij}')^{2}}{2\sigma^2}-\frac{1}{2}}}
\end{aligned}
\end{equation}

The overall probability-aware segmentation loss function is formulated as:
\begin{equation}
\label{eq16}
\mathcal{L}_{total}= \mathcal{L}_{Seg} + w\mathcal{L}_{KL}
\end{equation}
where \(w\) is a balance term to prevent \(\mathcal{L}_{KL}\) from dominating the update of parameters through backpropagation. Theoretically, the probability-informed segmentation loss function allows for a more nuanced model training that accounts for both the fidelity to the annotated data and the uncertainty inherent in pseudo labels, thus maintaining the integrity of the learning process even in less-than-ideal conditions.
\newpage
\subsection{Training and Inference Strategy}
\noindent \paragraph{Binary Classification}\quad Unlike binary segmentation map which is common in most deep learning tasks, the proposed probability-based pseudo label suffers from large data size and a single point has confidence scores for multiple organ classes, the sum of which might be greater than 1, which could be ambiguous. Thus, our model is trained to make inference of a single organ class, which is formulated as a binary classification task for each point.

\noindent \paragraph{Sampling of Dependency Scores}\quad During training, to accelerate the training process, we sample the dependency scores for only one time, while during inference, the dependency scores are sampled for $M$ times, and the final output is calculated as \cref{eq12}.

\section{Experiments}
\subsection{Implementation Details}
\label{app3}
Our model inherits the contracting-expanding pattern of UNETR~\cite{hatamizadeh2022unetr} but substitutes the encoder with a stack of Probabilistic Transformer blocks, each connected to the decoder through skip connections. We implemented our method using PyTorch~\cite{paszke2019pytorch} and MONAI\footnote{https://monai.io/}. All experiments were conducted on a single NVIDIA RTXA5000 GPU with 24GB of memory. We set the number of transformer encoders to 12 (L=12) with an embedding size of 768 (K=768). Each patch has a resolution of 16x16x16.

During training, we used the AdamW optimizer with an initial learning rate of 0.0001 and a batch size of 1, over 6,000 iterations. For inference, we employed a sliding window approach with a 50\% overlap. The number of sampled points for different labels is proportional to the volume of the corresponding organ: 200 points for the spleen, 400 points for the liver, and 50 points for each of the right and left kidneys.

We conducted experiments on two authoritative datasets: the BTCV dataset~\cite{landman2015miccai}, which includes multi-organ abdominal 3D CT scans acquired during the portal venous contrast phase, and the CHAOS dataset~\cite{kavur2021chaos}, involving the segmentation of four abdominal organs from MRI datasets acquired with two different sequences (T1-DUAL and T2-SPIR). We assessed the effectiveness of our methodology using two prevalent metrics: the DICE score, where higher scores indicate better performance, and the 95\% Hausdorff Distance (HD95), where lower values are preferable.

\subsection{Results} 
In this section, we present the results of our method, comparing it with leading pseudo label generation and fully supervised learning methods. Our approach shows superior performance, surpassing all other SOTA segmentation methods and even surpassing some fully supervised ones.

\paragraph{Comparison with state-of-the-art pseudo label generation methods}\quad \cref{tab:tab1} shows the quantitative results for four organs: spleen, liver, left kidney, and right kidney. We categorize the weakly supervised methods into two types of supervision: point-supervised learning and scribble-supervised learning. Point-supervised methods use a few annotated points to guide the segmentation, such as sparse~\cite{cciccek20163d}, convex~\cite{barber1996quickhull} and ours. ADNet~\cite{Hansen2022AnomalyDF} and ALPNet~\cite{ouyang2022self} are examples of using scribbles-supervised learning to generate pseudo labels. 

\begin{table}[htbp]
\centering
\caption{Comparison with SOTA weakly supervised methods on the BTCV and CHAOS datasets using the Dice metric. "LK" stands for "Left Kidney", and "RK" stands for "Right Kidney". Throughout this text, these abbreviations will consistently be used to refer to these anatomical structures.}
\vspace{0.1cm}
\label{tab:tab1}
\setlength{\tabcolsep}{2mm}{
    \begin{tabular}{lcccccc}
        \toprule
        Dataset & Method & Spleen & Liver & LK & RK \\
        \midrule
        \multirow{5}{*}{BTCV} 
        & Sparse~\cite{cciccek20163d} & 0.5515 & 0.4303 & 0.2532 & 0.2703 \\
        & Convex~\cite{barber1996quickhull} & 0.8232 & 0.6268 & 0.4037 & 0.3272 \\
        & ADNet~\cite{Hansen2022AnomalyDF} & 0.386 & 0.7389 & 0.1751 & 0.2382 \\
        & ALPNet~\cite{ouyang2022self} & 0.7455 & 0.7916 & 0.594 & 0.535 \\
        & \textbf{Ours} & \textbf{0.8279} & \textbf{0.8157} & \textbf{0.7599} & \textbf{0.7164} \\
        \midrule
        \multirow{5}{*}{CHAOS} 
        & Sparse~\cite{cciccek20163d} & 0.3693 & 0.5197 & 0.5675 & 0.559 \\
        & Convex~\cite{barber1996quickhull} & 0.7256 & 0.7659 & 0.564 & 0.7048 \\
        & ADNet~\cite{Hansen2022AnomalyDF} & 0.5641 & 0.7101 & 0.653 & 0.7652 \\
        & ALPNet~\cite{ouyang2022self} & 0.73 & 0.7036 & \textbf{0.7755} & 0.7706 \\
        & \textbf{Ours} & \textbf{0.7402} & \textbf{0.8205} & 0.6662 & \textbf{0.7716} \\
        \bottomrule
    \end{tabular}
}
\end{table}

From \cref{tab:tab1}, we can observe that our method achieves the best performance on both datasets, except for the left kidney on CHAOS dataset, where ALPNet is slightly better. Our method improves the Dice scores by up to 18.1\% and 10.2\% over the point-supervised methods, 58.4\% and 17.6\% over the scribble-supervised methods, and a large margin over the weakly supervised method on both datasets.

\begin{figure*}[htb]
  \centering
  \includegraphics[width=\linewidth]{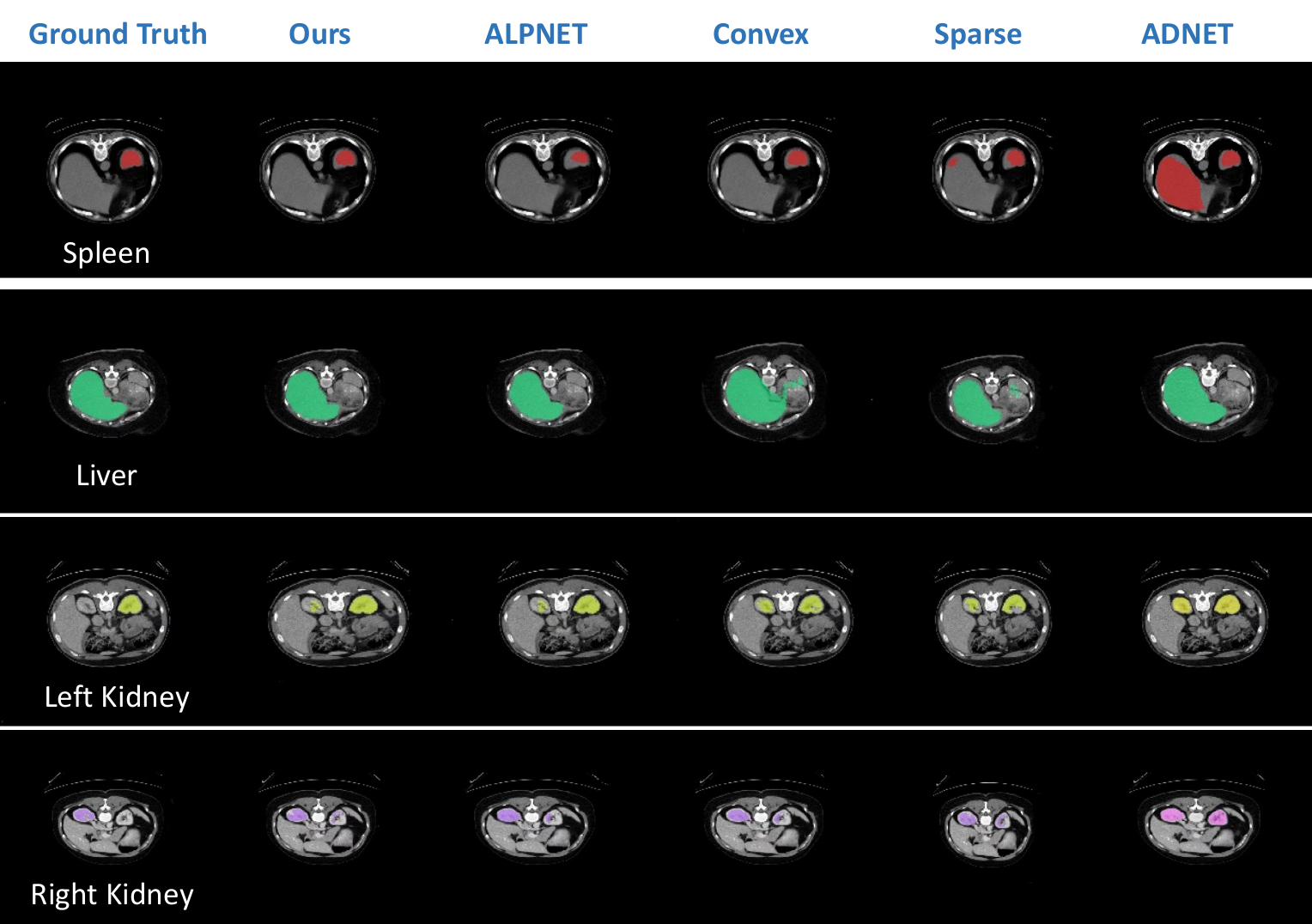}
  \caption{Qualitative comparison with weakly supervised methods.}
  \label{fig4}
  \vspace{-0.1in}
\end{figure*}

In conclusion, these results demonstrate the effectiveness of our method in producing high-quality segmentation results. The quantitative comparison in \cref{fig4} further highlights our method's proficiency in acquiring more accurate and comprehensive segments.

\paragraph{Comparison with SOTA fully supervised methods}\quad 
To highlight the efficacy of our proposed approach, we compare our weakly supervised method with state-of-the-art fully supervised methods on the BTCV dataset, including TransUnet \cite{chen2021transunet}, SwinUnet \cite{cao2021swin}, UCTransNet \cite{wang2022uctransnet} and UNETR \cite{hatamizadeh2022unetr}. It is paramount to note that this comparison is inherently imbalanced, as our method operates on notably sparser original annotations compared to the comprehensive annotations utilized by the aforementioned fully supervised methods.

\begin{table*}[ht]
\centering
\caption{Comparison with SOTA fully supervised methods on BTCV dataset using the Dice and HD95 metric.}
\vspace{0.1cm}
\label{tab:tab2}
\scalebox{0.86}{
    \setlength{\tabcolsep}{1.0mm}{
        \begin{tabular}{cccccccccc}
            \toprule
            \multicolumn{2}{c}{\multirow{2}{*}{Method}} & \multicolumn{2}{c}{Spleen} & \multicolumn{2}{c}{Liver} & \multicolumn{2}{c}{LK} & \multicolumn{2}{c}{RK}\\
            \cmidrule(r){3-4} \cmidrule(r){5-6} \cmidrule(r){7-8} \cmidrule(r){9-10}
            & & DICE↑ & HD95↓ & DICE↑ & HD95↓ & DICE↑ & HD95↓ & DICE↑ & HD95↓ \\
            \midrule
            \multirow{4}{*}{Fully} & TransUnet~\cite{chen2021transunet} & 0.8697 & 30.14 & 0.9341 & 10.21 & 0.7822 & 28.19 & 0.8431 & 29.24 \\
            & SwinUnet~\cite{cao2021swin} & 0.8294 & 27.38 & 0.9129 & 13.50 & 0.8017 & 63.74 & 0.8010 & 28.12 \\
            & UCTransNet~\cite{wang2022uctransnet} & 0.8176 & 29.22 & 0.8972 & 17.36 & 0.7822 & 22.77 & 0.7805 & 27.71 \\
            & UNETR~\cite{hatamizadeh2022unetr} & 0.9304 & 18.65 & 0.9017 & 39.26 & 0.9159 & 51.00 & 0.8945 & 6.35 \\
            \hline
            Weakly & Ours & \textbf{0.8279} & \textbf{63.09} & \textbf{0.8157} & \textbf{127.16} & \textbf{0.7599} & \textbf{135.88} & \textbf{0.7164} & \textbf{116.22} \\
            \bottomrule
        \end{tabular}
    }
}
\end{table*}

Despite this inherent disparity, as delineated in \cref{tab:tab2}, our method exhibits performances that are remarkably on par with, and in certain metrics, even surpass, those achieved by fully supervised counterparts. For instance, our method eclipses UCTransNet in spleen segmentation, showcasing the distinct advantages of our probabilistic weakly supervised approach.

\begin{figure}[htb]
  \centering
  \includegraphics[width=0.9\linewidth]{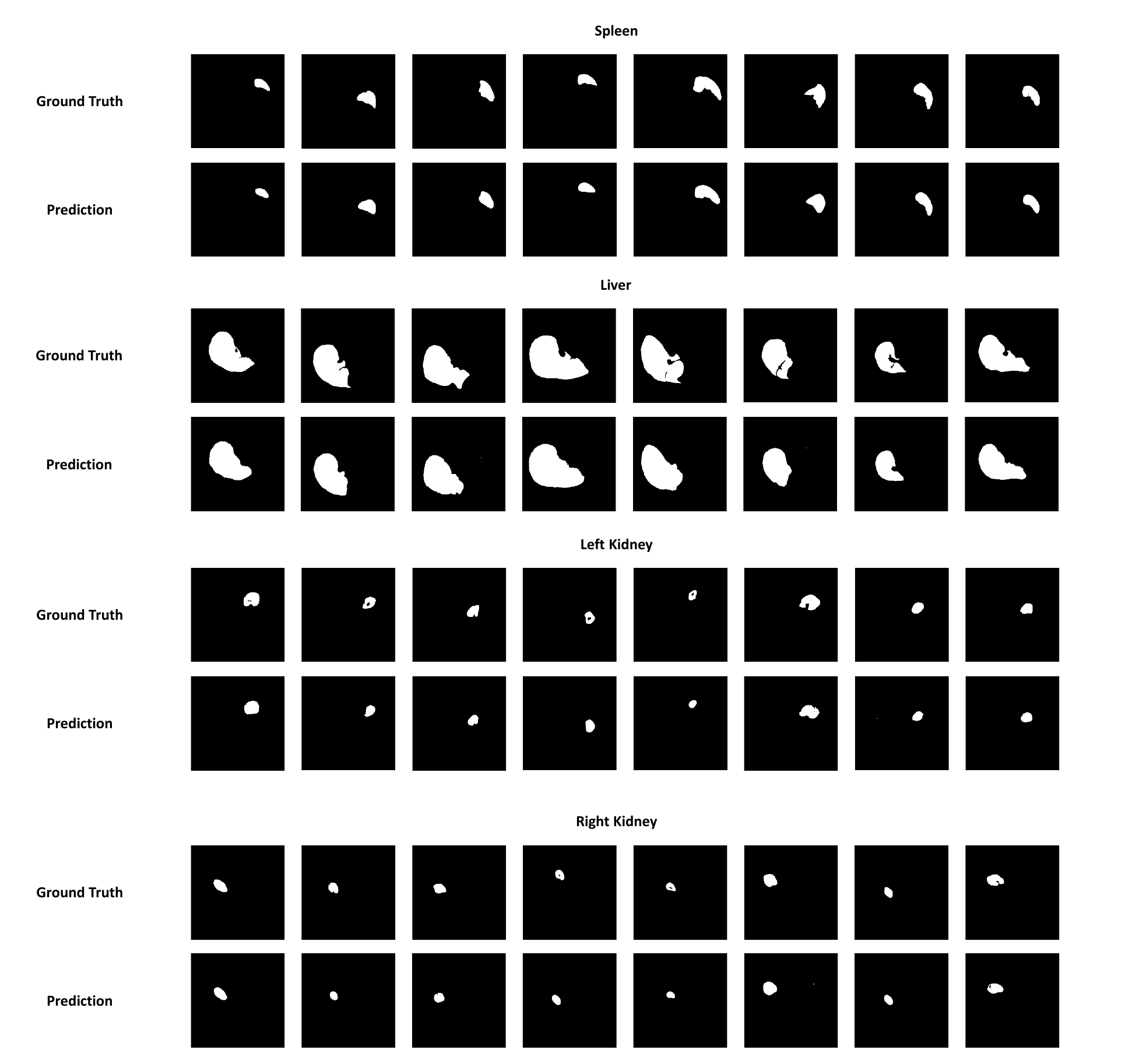}
  \caption{Qualitative results of segmentation prediction.} 
  \label{fig:QualitativeResults}
\end{figure}

To provide a more intuitive understanding of our method's performance, we present qualitative segmentation results in \cref{fig:QualitativeResults}, which demonstrates our model's capability to accurately delineate organ boundaries across different anatomical structures. The visual results confirm that our weakly supervised approach can achieve segmentation quality comparable to fully supervised methods, while requiring significantly less annotation effort. 

\begin{figure}[htb]
  \centering
  \includegraphics[width=\linewidth]{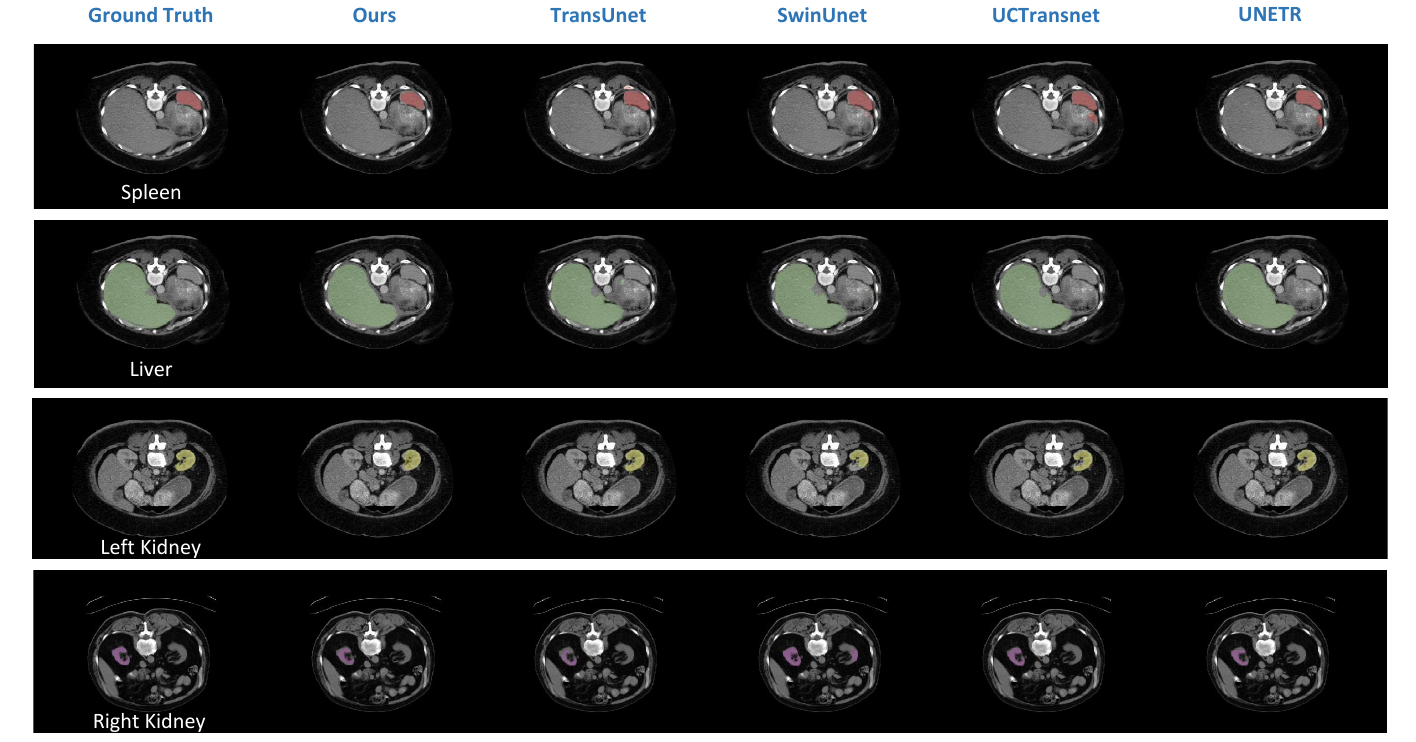}
  \caption{Qualitative comparison with fully supervised methods.}
  \label{figx}
  \vspace{0.2in}
\end{figure}

In \cref{figx}, we provide visual comparisons of our method against several fully supervised approaches, demonstrating its effectiveness in accurately segmenting key regions of interest. Despite limited supervision during training, our approach achieves performance comparable to fully supervised methods, underscoring its potential as a strong alternative.

In conclusion, our method demonstrates its prowess and superior adaptability, ensuring commendable accuracy even with limited annotations and emphasizing its potential as a robust solution in the realm of medical image segmentation.

\subsection{Ablation Study}
\label{ablation}
In the ablation study section, we investigate the integration of a probabilistic mechanism across three key aspects of our framework: pseudo-label generation, network structure, and loss function. This section also covers additional ablation studies exploring parameters like sampled points and variance selection, offering insights into their impact on our pipeline's performance.

\paragraph{Effectiveness of Probability-based Pseudo Label Generation} \quad
We assess the performance of a probabilistic mechanism using a semi-supervised technique with random point selection at various thresholds, aiming to mimic real-world scenarios with irregular point distributions or feature absence. Our goal is to ascertain if this approach yields consistent results across typical real-world data distributions, bridging the gap between laboratory and real-life settings, and ensuring effectiveness in both controlled and varied authentic environments. 

\begin{table*}[htb]
\centering
\caption[Performance assessed using a semi-supervised probabilistic approach]%
{Performance assessed using a semi-supervised probabilistic approach with random point selection at various thresholds, simulating real-world scenarios with irregular point distributions. The study evaluates the efficacy of probability-based pseudo label generation, ensuring uniform performance across common data distributions while smoothly bridging the gap between laboratory insights and real-world implementations.}
\vspace{0.1cm}
\label{tab:ppgx}
\scalebox{0.88}{
    \setlength{\tabcolsep}{1.5mm}{
        \begin{tabular}{l|c|cccc|cccc}
            \toprule
            \multirow{2}{*}{Dataset} & \multirow{2}{*}{Ratio} & \multicolumn{4}{c|}{Dice↑} & \multicolumn{4}{c}{HD95↓} \\
            & & Spleen & Liver & LK & RK & Spleen & Liver & LK & RK \\
            \midrule
            \multirow{6}{*}{BTCV} & 10\% & 0.5425 & 0.6996 & 0.5797 & 0.5439 & 369.62 & 310.38 & 148.71 & 122.25 \\
            & 30\% & 0.5747 & 0.8136 & 0.5784 & 0.5259 & 388.08 & 133.17 & 297.53 & 202.54 \\
            & 50\% & 0.6178 & 0.7845 & 0.3971 & 0.3894 & 352.41 & 281.86 & 347.38 & 325.83 \\
            & 70\% & 0.5961 & 0.7715 & 0.5686 & 0.3799 & 354.61 & 308.32 & 326.35 & 341.95 \\
            & 90\% & 0.6652 & 0.8060 & 0.3927 & 0.3889 & 172.67 & 144.89 & 320.72 & 324.61 \\
            & Ours & \textbf{0.8279} & \textbf{0.8157} & \textbf{0.7599} & \textbf{0.7164} & \textbf{63.09} & \textbf{127.16} & \textbf{135.88} & \textbf{116.22} \\
            \midrule
            \multirow{6}{*}{CHAOS} & 10\% & 0.3803 & 0.6544 & 0.4564 & 0.4110 & \textbf{36.40} & 49.93 & \textbf{51.88} & 36.50 \\
            & 30\% & 0.4179 & 0.7199 & 0.5858 & 0.6125 & 79.14 & 51.06 & 132.24 & 41.71 \\
            & 50\% & 0.4054 & 0.6504 & 0.5819 & 0.5916 & 32.91 & 49.03 & 101.22 & 62.44 \\
            & 70\% & 0.4903 & 0.7390 & 0.5851 & 0.6200 & 56.04 & 61.85 & 103.40 & 52.12 \\
            & 90\% & 0.5455 & 0.7450 & 0.6287 & 0.6299 & 53.55 & 61.06 & 101.05 & 42.06 \\
            & Ours & \textbf{0.7402} & \textbf{0.8205} & \textbf{0.6662} & \textbf{0.7716} & 53.11 & \textbf{48.75} & 93.48 & \textbf{36.01} \\
            \bottomrule
        \end{tabular}
    }
}
\end{table*}

\cref{tab:ppgx} shows marked improvements in segmentation accuracy, evident from significant Dice Score increases, such as from 0.5425 to 0.8279 for the spleen in the BTCV dataset and from 0.6544 to 0.8205 for the liver in the CHAOS dataset. The HD95 metrics also improved, although they are sensitive to extreme cases, particularly in complex anatomical regions like the left kidney in the CHAOS dataset. This sensitivity is a common issue for weakly supervised methods and is not unique to our approach.

These results demonstrate the method's adaptability to real-world irregularities and its robustness across different clinical scenarios. The consistent performance across various organs and datasets proves its real-world applicability, narrowing the gap between lab and real-life settings. Additionally, the enhanced segmentation accuracy has important clinical implications, affecting clinical decisions and patient care. Our experiments highlight our method's technical and clinical potential, suggesting it for widespread use due to its reliability in diverse conditions.

\begin{figure*}[htb]
  \centering
  \includegraphics[width=\linewidth]{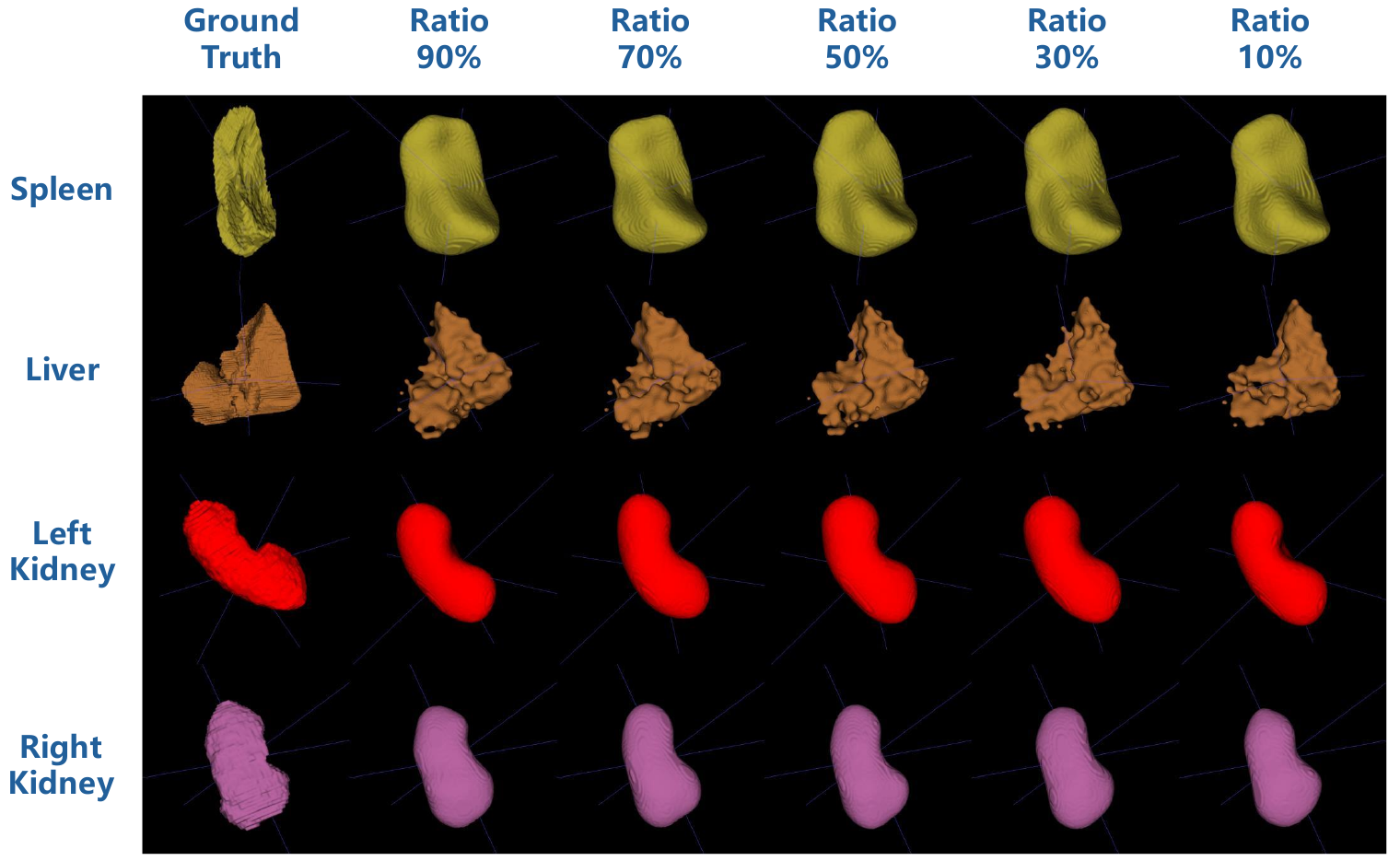}
  \caption{Visualization results on BTCV dataset showing the comparison of pseudo-labels generated under different thresholds of random original labels.}
  \label{fig:btcv}
  \vspace{-0.1in}
\end{figure*}

\begin{figure*}[htb]
  \centering
  \includegraphics[width=\linewidth]{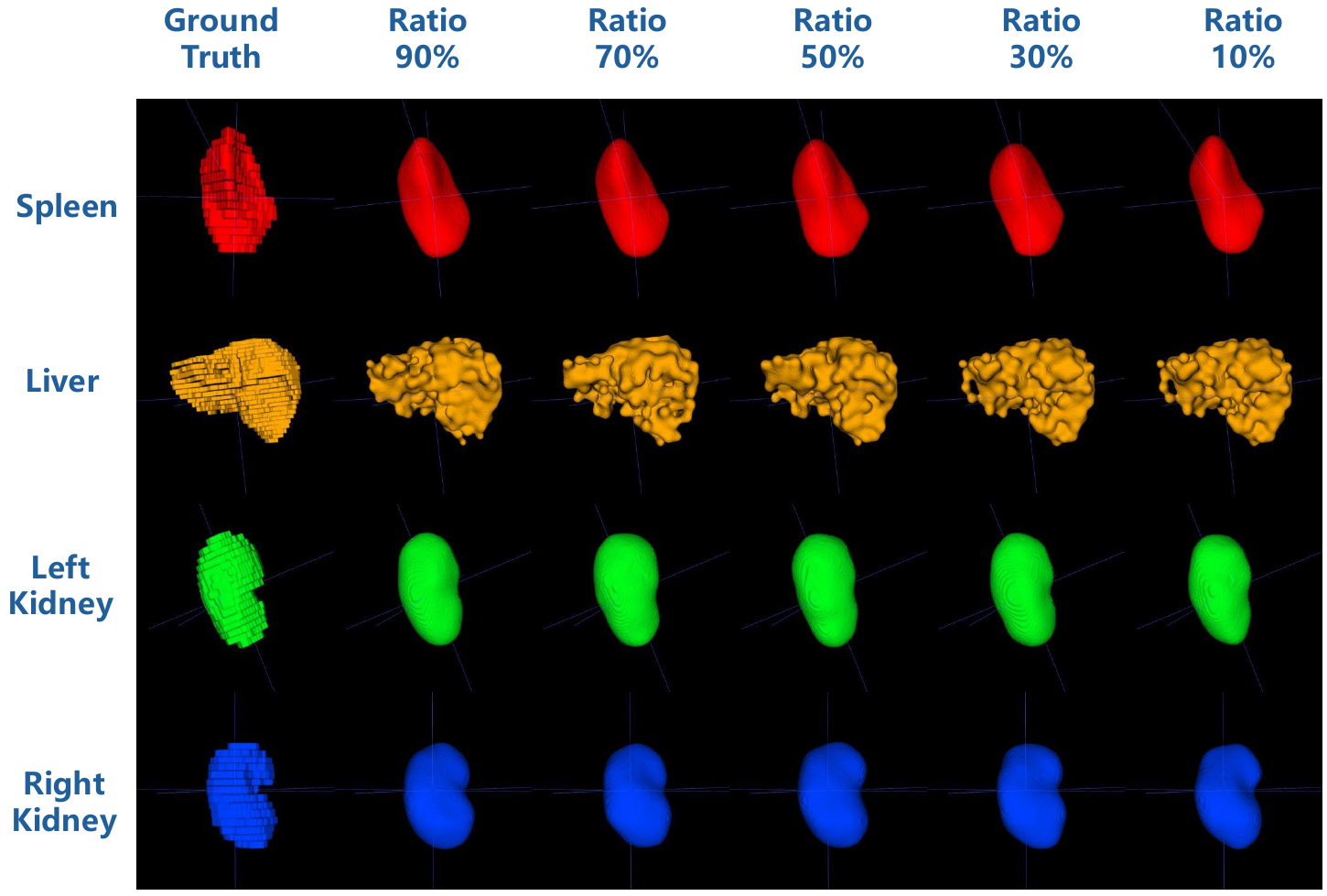}
  \caption{Visualization results on CHAOS dataset demonstrating the effectiveness of our method across different threshold settings.}
  \label{fig:chaos}
  \vspace{-0.1in}
\end{figure*}

\cref{fig:btcv} and \cref{fig:chaos} show a comparison of the visualizations of pseudo-labels generated by our method under different thresholds of random original labels in two multi-modal datasets. Since our method is based on sparse label generation, the pseudo-labels produced by our approach tend to be much smoother compared to the ground truth. This is a shortcoming of our method, as well as a challenge that needs to be addressed in the field of weak supervision. Consequently, the boundary information is slightly inferior, which also explains the slightly higher HD95 distance. However, our method can still produce pseudo-labels with an overall contour even when sampling from randomly labeled points with very low thresholds. This makes our method more advantageous for datasets with very few or poorly labeled annotations.

\begin{table*}[htb]
\centering
\caption{Quantitative results of ablation study for network structure and loss function. For these parts of comparison, only the methods corresponding to that part are varied, while the methods in the other parts are kept constant. ''Ours'' column represents the results of our complete approach.}
\vspace{0.1cm}
\label{tab:merged_table}
\scalebox{0.9}{
    \setlength{\tabcolsep}{1.5mm}{
        \begin{tabular}{lcccccccccc}
            \toprule
            \multirow{2}{*}{Dataset} & \multirow{2}{*}{Metric} & \multirow{2}{*}{Organ} & \multicolumn{2}{c}{Network} & \multicolumn{4}{c}{Loss Function} & \multirow{2}{*}{Ours} \\
             &  &  & MSA & SA & DICE & CE & DCE & Focal &  \\
            \midrule
            \multirow{8}{*}{BTCV} 
            & \multirow{4}{*}{DICE↑} & Spleen & 0.817 & 0.6013 & 0.3561 & 0.6341 & 0.7665 & 0.7853 & \textbf{0.8279} \\
            &  & Liver & 0.7719 & 0.7865 & 0.5992 & 0.7767 & 0.7753 & 0.4025 & \textbf{0.8157} \\
            &  & LK & 0.4252 & 0.4207 & 0.2599 & 0.3963 & 0.6112 & 0.5653 & \textbf{0.7599} \\
            &  & RK & 0.5445 & 0.354 & 0.3403 & 0.4839 & 0.506 & 0.3675 & \textbf{0.7164} \\
            \cline{2-10}
            & \multirow{4}{*}{HD95↓} & Spleen & 285.41 & 385.36 & 373.83 & 362.6 & 316.56 & 350.44 & \textbf{63.09} \\
            &  & Liver & 306.89 & 295.71 & 321.67 & 300.85 & \textbf{82.13} & 344.48 & 127.16 \\
            &  & LK & 330.94 & 323.53 & 371.04 & 341.61 & \textbf{95.04} & 107.78 & 135.88 \\
            &  & RK & 135.78 & 325.88 & 342.03 & 318.26 & 120.08 & 216.56 & \textbf{116.22} \\
            \midrule
            \multirow{8}{*}{CHAOS} 
            & \multirow{4}{*}{DICE↑} & Spleen & 0.7145 & 0.7481 & 0.2999 & 0.5058 & 0.4447 & 0.3442 & \textbf{0.7402} \\
            &  & Liver & 0.7542 & 0.7781 & 0.6596 & 0.6092 & 0.7033 & 0.379 & \textbf{0.8205} \\
            &  & LK & 0.6279 & 0.6485 & 0.3537 & 0.5037 & 0.446 & 0.6297 & \textbf{0.6662} \\
            &  & RK & 0.6284 & 0.6716 & 0.5952 & 0.6514 & 0.6926 & 0.5693 & \textbf{0.7716} \\
            \cline{2-10}
            & \multirow{4}{*}{HD95↓} & Spleen & 74.09 & 164.91 & 189.34 & 122.37 & 53.44 & 74.93 & \textbf{53.11} \\
            &  & Liver & 93.52 & 75.39 & 180.32 & 45.81 & \textbf{32.02} & 76.86 & 48.75 \\
            &  & LK & 84.26 & 106.01 & 133.96 & 106.79 & 46.39 & \textbf{31.59} & 93.48 \\
            &  & RK & 114.14 & 98.64 & 168.17 & 57.92 & 37.75 & 56.69 & \textbf{36.01} \\
            \bottomrule
        \end{tabular}
    }
}
\vspace{0.1in}
\end{table*}

\paragraph{Effectiveness of Probabilistic Transformer Network Structure} \quad
We investigate the impact of the probabilistic mechanism in the network architecture. \cref{tab:merged_table} presents our experimental results, comparing the performance of the Self-Attention (SA) and the Multi-head Self-Attention (MSA) methods. These results indicate the heightened accuracy and reliability of PMSA in producing segmentation results that closely align with the actual anatomical structures, demonstrating the significance of considering probabilistic modeling in our transformer network.

\paragraph{Effectiveness of Probability-informed Segmentation Loss Function} \quad
We examine the effectiveness of our designed loss function, as presented in \cref{tab:merged_table}. The conclusive results underscore the outstanding efficacy of our approach. Our approach consistently achieves higher effectiveness scores, demonstrating its ability to deliver more accurate and coherent segments. Compared to our approach, the existing non-probabilistic loss functions, specifically DICE, Cross-Entropy (CE), combined Dice-Cross-Entropy (DCE), and Focal demonstrate suboptimal performance, especially in segmenting the liver and both kidneys. These findings underscore the limitations of the existing loss functions and underscore the superiority of our designed probability-informed loss function in achieving improved 3D medical image segmentation results.

\subsection{Parameters Exploration}
In our study, we conduct in-depth ablation analyses on two crucial parameters. Specifically, the number of sampled points, as detailed in our annotation strategy (\cref{ppmg}), plays a pivotal role in pseudo label generation. Additionally, the selection of variance $\sigma^2$ in computing the KL loss, a critical hyperparameter, is meticulously evaluated to determine its influence on segmentation accuracy. 

\begin{table*}[htb]
\centering
\caption{Illustration of the number of sampled points.}
\vspace{0.1cm}
\label{tab:n}
\scalebox{0.97}{
    \setlength{\tabcolsep}{1.3mm}{
        \begin{tabular}{l|c|cccc|cccc}
            \toprule
            \multirow{2}{*}{Dataset} & \multirow{2}{*}{n} & \multicolumn{4}{c|}{Dice↑} & \multicolumn{4}{c}{HD95↓} \\
            & & Spleen & Liver & LK & RK & Spleen & Liver & LK & RK \\
            \midrule
            \multirow{4}{*}{BTCV} & 50 & 0.6392 & 0.1081 & 0.6276 & 0.5678 & 336.25 & 127.16 & 135.88 & 292.41 \\
            & 100 & 0.8001 & 0.7307 & 0.6366 & 0.4772 & 174.41 & 295.54 & 197.66 & 312.45 \\
            & 150 & 0.5462 & 0.7164 & 0.3856 & 0.5756 & 348.11 & 304.69 & 333.50 & 183.65 \\
            & 200 & 0.8279 & 0.8157 & 0.7599 & 0.7164 & 63.08 & 265.79 & 266.17 & 116.22 \\
            \midrule
            \multirow{4}{*}{CHAOS} & 200 & 0.7356 & 0.8205 & 0.6662 & 0.6530 & 77.31 & 74.05 & 93.48 & 94.50 \\
            & 250 & 0.6982 & 0.7698 & 0.5687 & 0.7716 & 148.26 & 103.01 & 107.56 & 36.01 \\
            & 300 & 0.7402 & 0.7970 & 0.5978 & 0.6667 & 130.12 & 48.79 & 103.09 & 112.75 \\
            & 350 & 0.6711 & 0.7872 & 0.5516 & 0.6627 & 53.11 & 48.75 & 101.34 & 99.07 \\
            \bottomrule
        \end{tabular}
    }
}
\end{table*}

\paragraph{Impact of the Number of Sampled Points} \quad As shown in \cref{tab:n}, we conducted a detailed investigation into the impact of the number of sampled points on segmentation performance. This experiment was designed to keep all other parameters constant, varying only the number of sampled points used in pseudo label generation. The results from this comparative study provide intriguing insights into the optimal balancing of sampled points for effective segmentation.

A key observation from the BTCV and CHAOS datasets is the non-linear relationship between the number of sampled points and the segmentation performance. Specifically, we noticed that both extremely low and high numbers of sampled points do not necessarily yield the best segmentation results. For instance, in the BTCV dataset, a sample size of 50 points resulted in suboptimal Dice Scores and HD95 metrics across all organs, suggesting inadequate coverage of the organ's semantic space. Conversely, at 200 points, while some organs like the spleen and liver showed marked improvements in Dice Scores and reduced HD95 values, indicating better segmentation, others like the left kidney did not show a consistent pattern of improvement.

This phenomenon can be attributed to the fact that a very low number of points may fail to provide sufficient information to cover the entire organ, leading to poor segmentation performance. On the other hand, a very high number of points could introduce noise or outliers, potentially hampering the segmentation accuracy. These additional points, rather than contributing useful information, might act as anomalies, detracting from the model's ability to accurately delineate organ boundaries.

Our results highlight the importance of an optimal range of sampled points in our probabilistic pseudo label generation, striking a balance between comprehensive feature representation and minimizing noise. This balance is crucial for enhancing segmentation accuracy while efficiently utilizing limited annotation resources, proving especially beneficial in scenarios where full supervision is not feasible. The findings underscore the significance of carefully selecting the number of sampled points to achieve effective annotation efficiency and robust segmentation outcomes.

\begin{table*}[htb]
\centering
\caption{Illustration of the number of selection of variance.}
\vspace{0.1cm}
\label{tab:sigma}
\scalebox{0.97}{
    \setlength{\tabcolsep}{1.3mm}{
        \begin{tabular}{l|c|cccc|cccc}
            \toprule
            \multirow{2}{*}{Dataset} & \multirow{2}{*}{\textbf{\(\sigma^2\)}} & \multicolumn{4}{c|}{Dice Score↑} & \multicolumn{4}{c}{HD95↓} \\
            & & Spleen & Liver & LK & RK & Spleen & Liver & LK & RK \\
            \midrule
            \multirow{3}{*}{BTCV} & 0.1 & 0.5104 & 0.7478 & 0.5474 & 0.5192 & 394.87 & 299.2 & 329.39 & 273.61 \\
            & 1 & 0.8279 & 0.8157 & 0.7599 & 0.7164 & 63.09 & 265.79 & 266.17 & 116.22 \\
            & 10 & 0.5754 & 0.7691 & 0.6103 & 0.3468 & 388.11 & 312.36 & 238.93 & 323.79 \\
            \midrule
            \multirow{3}{*}{CHAOS} & 0.1 & 0.5472 & 0.5707 & 0.3669 & 0.3813 & 56.26 & 51.06 & 31.72 & 56.14 \\
            & 1 & 0.7402 & 0.8205 & 0.6662 & 0.7716 & 53.11 & 48.75 & 93.48 & 36.01 \\
            & 10 & 0.5399 & 0.5413 & 0.4058 & 0.6664 & 164.89 & 45.93 & 68.55 & 112.75 \\
            \bottomrule
        \end{tabular}
    }
}
\end{table*}

\paragraph{Comparison of the Selection of Variance} 
When calculating the KL loss in our probability-informed segmentation loss function, the variance $\sigma^2$ serves as a hyperparameter that needs to be manually determined. To ensure experimental rigor, we investigate the effects of different variances on segmentation accuracy. \cref{tab:sigma} illustrates that the choice of variance in the KL loss significantly influences the final results. We observe that when setting the variance to 1, our model achieves the highest DICE score and the lowest HD95 value. Based on these empirical findings, we establish $\sigma^2$ as 1 in our method. By conducting this analysis, we enhance the reliability of our experimental setup and demonstrate the importance of selecting an appropriate variance for the KL loss. The chosen value of $\sigma^2$ contributes to optimizing the segmentation performance and ensures the robustness of our method.

\section{Conclusion}
In this work, we present a novel probability-based framework for 3D medical image segmentation under weak supervision, showing marked accuracy improvements over state-of-the-art methods. This approach not only pioneers new and efficient segmentation strategies but also ensures precision with minimal annotations, promising significant real-world applicability.

\section{Acknowledgment}
This work was supported in part by U.S. NIH grants R01GM134020 and P41GM103712, NSF grants DBI-1949629, DBI-2238093, IIS-2007595, IIS-2211597, and MCB-2205148. This work was supported in part by UPMC Enterprises, Oracle Cloud credits and related resources provided by Oracle for Research, and the computational resources support from AMD HPC Fund.

\clearpage
\bibliographystyle{elsarticle-num} 
\bibliography{main.bib}

\clearpage
\appendix
\begin{appendix}

\section{Sampling of the Dependency Scores}
\label{app2}
The dependency scores of a deeper PMSA layer are mutually independent but only rely on the former layers.
Therefore, we have:
\begin{equation}
    \alpha_{ij} \sim \mathcal{N}(\mu_{ij}, \sigma^{2}_{ij}),
\end{equation}
\begin{equation}
\label{depend}
   \alpha_{l}\sim p(\alpha_{l}|X',\Theta,\alpha_{l-1}...,\alpha_{1}),\quad l=1,...,L.
\end{equation}
where the mean $\mu_{ij}$ and the variance $\sigma^{2}_{ij}$ are calculated with $q_{i}$ and $k_{j}$ using a multilayer perceptron (MLP), and $\alpha_{l}$ denotes the dependency scores of the PMSA layer of the $l$th transformer block. $\Theta$ denotes all the deterministic parameters in the model.

In this way, given the input image $X'$, the distribution of the output segmentation map $y'$ is calculated as:
\begin{equation}
\begin{split}
 P(y'|X', \Theta)=E_{\alpha \sim p(\alpha|X',\Theta)}[P(y'|X',\Theta,\alpha)]\\
 =\int_{\alpha}{P(y'|X',\Theta,\alpha)p(\alpha|X',\Theta)d\alpha}.
\end{split}
\end{equation}

During inference, to approximate the integral of \cref{eq11}, we sample all the dependency scores independently for $M$ times and calculate the final segmentation output $y^{*}$ where $\alpha_{m}$ denotes the sampled dependency scores:
 
\begin{equation} 
\label{appro}
y^{*} = \mathop{\argmax}_{y'} \sum\limits_{m=1}^{M} \frac{1}{M} P(y'|X',\Theta,\alpha_{m})
\end{equation}
Given that the dependency scores within the same PMSA layer are independent of each other, and the dependency scores of deeper PMSA layer are dependent on those of former PMSA layers, as indicated by \cref{depend}, \cref{eq11} could be written as:

\begin{equation}
\begin{aligned}
     \label{integ} &P(y'|X', \Theta)= \int_{\alpha} P(y'|X',\Theta,\alpha)p(\alpha|X',\Theta) \, d\alpha \\
     &= \int_{\alpha_{1}} \cdots \int_{\alpha_{L}} P(y'|X',\Theta,\alpha_{1},\ldots,\alpha_{L}) \times \\
     &\quad p(\alpha_{L} |X',\Theta,\alpha_{1},\ldots,\alpha_{L-1}) \, d\alpha_{L} \cdots p(\alpha_{1}|X',\Theta) \, d\alpha_{1} \\
     &\approx \int_{\alpha_{1}} \cdots \int_{\alpha_{L-1}} \frac{1}{M_{L}} \sum\limits_{m_{L}=1}^{M_{L}} P(y'|X',\Theta,\alpha_{1},\ldots,\alpha_{L_{m_{L}}}) \times \\
     &\quad p(\alpha_{L-1}|X',\Theta,\alpha_{1},\ldots,\alpha_{L-2}) \, d\alpha_{L-1} \cdots p(\alpha_{1}|X',\Theta) \, d\alpha_{1} \\
     &\approx \frac{1}{M_{1}} \sum\limits_{m_{1}=1}^{M_{1}} \cdots \frac{1}{M_{L}} \sum\limits_{m_{L}=1}^{M_{L}} P(y'|X',\Theta,\alpha_{1_{m_{1}}},\ldots,\alpha_{L_{m_{L}}}) \\
     &\approx \frac{1}{M} \sum\limits_{m_{1}=1}^{M_{1}} \cdots \sum\limits_{m_{L}=1}^{M_{L}} P(y'|X',\Theta,\alpha_{1_{m_{1}}},\ldots,\alpha_{L_{m_{L}}}) \\
     &\approx \frac{1}{M} \sum\limits_{m=1}^{M} P(y'|X',\Theta,\alpha_{1_{m}},\ldots,\alpha_{L_{m}})
\end{aligned}
\end{equation}
where \(M=\prod \limits_{l=1}^{L} M_{l}\), which we empirically set as 6 in our experiments.

\end{appendix}
\end{document}